\documentclass[
reprint,
superscriptaddress,
 aps,
]{revtex4-2}
\usepackage{amsmath,amssymb,mathrsfs,multirow}
\usepackage{graphicx}
\usepackage{dcolumn}
\usepackage{bm,slashed}
\usepackage[colorlinks=true,linkcolor=blue,urlcolor=blue,citecolor=blue]{hyperref}
\usepackage[compat=1.1.0]{tikz-feynman}
\usepackage{amsmath}
\usepackage[utf8]{inputenc}
\usepackage{slashed}
\usepackage{subfig}
\usepackage{graphicx}
\usepackage{float}
\usepackage{multirow}

\begin{document}

\title{\boldmath Probing $\Delta L=2$ lepton number violating SMEFT operators at the same-sign muon collider}

\author{Subhaditya Bhattacharya}
\email{subhab@iitg.ac.in}
\affiliation{Department of Physics, Indian Institute of Technology Guwahati, North Guwahati, 781039, India}
\author{Soumyajit Datta}
\email{soumyajitdattaphys@gmail.com}
\affiliation{Department of Physical Sciences, Indian Institute of Science Education and Research Kolkata, Mohanpur, 741246, India}
\author{Abhik Sarkar}
\email{sarkar.abhik@iitg.ac.in}
\affiliation{Department of Physics, Indian Institute of Technology Guwahati, North Guwahati, 781039, India}

\begin{abstract}
We investigate lepton number violation (LNV) induced by $\Delta L = 2$ dimension-seven Standard Model Effective Field Theory (SMEFT) operators in the context of same-sign muon colliders. Specifically, we study $\mu^+ \mu^+ \rightarrow W^+W^+/\;W^+qq'$ production at the $\mu$TRISTAN at $\sqrt{s} =$ 2 TeV with an integrated luminosity of 1 ab$^{-1}$, using the final-state signature comprising two fat jets. This process is sensitive to eight distinct SMEFT operators, providing a unique avenue for testing LNV beyond the Standard Model. We determine the maximal sensitivity to these operators and compare our results with existing LHC constraints and future projections from the FCC. Our findings highlight the potential of same sign muon colliders to serve as powerful probes of LNV and New Physics (NP) at the TeV scale.
\end{abstract}

\maketitle


\section{Introduction}
\label{sec:introduction}
Symmetries play an important role in describing fundamental particles and interactions. Lepton number symmetry is one such accidental global symmetry. 
Following the observation that charge and neutral current weak (as well as electromagnetic) interactions preserve such symmetry, 
in the Standard Model (SM) of particle physics, lepton number (more specifically, electron number, muon number and tau number) 
is a strictly conserved quantity. This conservation is ensured by the field definitions in conjunction with the gauge charges 
assigned to them, and there are no renormalizable terms that would allow lepton number violation (LNV). However, neutrino oscillations \cite{Pontecorvo:1957cp} demonstrate that neutrinos have mass, which the SM alone cannot explain. 
If we assume the presence of right handed neutrinos, which are iso-singlets with zero hypercharge, then a possible Majorana mass term can lead to LNV, 
which would show up in certain processes, such as neutrinoless double-beta decay \cite{PAS1999194,PAS200135,Deppisch:2012nb,Deppisch:2017ecm,Cirigliano:2017djv,deGouvea:2007qla,Ali:2007ec}. Non-observation of such processes places stringent constraints on LNV \cite{KamLAND-Zen:2022tow,Majorana:2022udl,CUPID-Mo:2022cel,NEXT:2023daz}, and set strong upper bounds on such processes, 
implying that if LNV exists, it must be either highly suppressed or occur at energy scales beyond current experimental reach. We shall note that lepton flavor symmetry in the SM is broken by the Yukawa terms yielding different masses to different lepton generations. Therefore, the constraints from neutrinoless double beta decay, that affect the operators with electron flavor, do not necessarily limit scenarios that involve heavier leptons. In this study, we focus on LNV interactions associated with the muon flavor.

LNV has been extensively studied in the literature, particularly in the context of neutrino mass generation mechanisms such as the Seesaw mechanism \cite{MINKOWSKI1977421, Gell-Mann:1979vob, PhysRevD.22.2227, PhysRevLett.44.912, PhysRevD.25.774}, radiative models \cite{Zee:1980ai,Babu:2002uu,Krauss:2002px,Ma:2006km,Cai:2017jrq}, and related frameworks. As we don't know about the nature of New Physics (NP), the Standard Model Effective Field Theory (SMEFT) \cite{Brivio:2017vri} serves as the default framework to study physics beyond the SM. It encapsulates NP effects through higher-dimensional operators built from SM fields while preserving the gauge symmetry of the SM. The SMEFT Lagrangian can be written as
\begin{equation}
    \mathcal{L}_{\rm SMEFT} = \mathcal{L}_{\rm SM} + \sum_{i} \,\frac{C^{(d)}_{i} \mathcal{O}^{(d)}_{i}}{\Lambda^{d-4}} \,,
\end{equation}
where, $\mathcal{O}^{(d)}_i$ are the operators of mass dimension $d$; where $\Lambda$ is the effective NP scale and $C^{(d)}_{i}$ are the dimensionless Wilson coefficients. The lepton number symmetry is explicitly violated at the leading order extension of the Standard Model (SM) by the dimension-5 Weinberg operator ($\Delta L = 2$) \cite{Weinberg:1979sa}: 
\begin{equation}
        (C_{\mathcal{W}}/\Lambda)\; \epsilon_{ij} \epsilon_{mn}\left(\overline{L^{c}_{i}}L_{m}\right)H_{j}H_{n},
\end{equation}
where $\epsilon$ indicates completely antisymmetric tensor of rank 2, and $H$ stands for SM Higgs isodoublet. Models like Type-I, Type-II or Type-III Seesaw frameworks
can generate such operators where the heavy physics is integrated out. Owing to its structure, the Weinberg operator involves only SM neutrinos and the Higgs field, leading to a Majorana mass term for neutrinos after electroweak symmetry breaking (EWSB). {However, the direct neutrino mass measurement (electron-based) \cite{KATRIN:2021uub} requires the NP scale associated with electron flavour $\Lambda_{ee} \sim 10^{12}$ TeV (assuming $C_{\mathcal{W}} \sim 1$), which lies far beyond the reach of current experimental facilities. This stringent constraint on the Weinberg operator arises under the assumption that it solely accounts for neutrino mass generation and is dependent on the flavor composition of the operator; while relaxing these assumptions significantly weakens the bound. The Weinberg operator has been extensively studied in the literature \cite{CMS:2022hvh,Fuks:2020zbm}, a discussion on which is provided in Appendix \ref{sec:appA}.} Within the SMEFT framework, baryon number violation first appears at dimension 6 level through four operators that simultaneously violate both baryon and lepton number ($\Delta B = \Delta L = 1$). Each of these operators involves three quark fields and one lepton field \cite{Grzadkowski:2010es}. At dimension-7, there are 18 independent operators (excluding redundancies), all of which violate the lepton number: 12 of the $\Delta L = 2$ type and 6 which violate both baryon and lepton number simultaneously ($\Delta B = \Delta L = 1$) \cite{Lehman:2014,Liao:2016hru,delAguila:2012nu,Bhattacharya:2015vja}.

In this work, we focus on the $\Delta L = 2$ operators, which can be effectively probed at future same-sign lepton colliders such as the $\mu$TRISTAN \cite{Hamada:2022mua,Heusch:1995yw}. 
At the same-sign muon collider $\mu$TRISTAN, the initial state itself carries a net lepton number ($\Delta L = 2$), making it ideal for studying lepton number violating processes, which are characterized by a final state with zero total lepton number. We should note that as neutrinos escape detection, their presence in the final state may very well be within the lepton number conserving paradigm, but, can potentially disguise the zero lepton final state signal. Missing energy turns very handy in discriminating such cases, as we elaborate.
$\Delta L = 2$ operators have previously been studied in the context of the LHC \cite{Fridell:2023rtr,Bhattacharya:2015vja} where same sign dilepton final state provides the 
main signal; however, huge $t\bar{t}$ background can camouflage such final states with same sign leptons arising from $b \to c \ell \nu_\ell$ and leptonic decays of $W$. Similar studies have also been done in the context of $\mu^{+}\mu^{-}$ colliders \cite{Frigerio:2024jlh}. Same-sign lepton colliders on the contrary, are expected to offer a cleaner and more sensitive environment for their exploration. There have been a number of model-specific studies concerning LNV at the $\mu$TRISTAN \cite{deLima:2024ohf,Jiang:2023mte,Santiago:2024zpc,Das:2024kyk}; while our study focuses specifically on muon-flavored $\Delta L = 2$ operators contributing via: $\mu^{+} \mu^{+} \rightarrow W^{+}W^{+}/\,W^{+}qq'$ at the $\mu$TRISTAN, where we show how the sensitivity can be enhanced.

{From the collider perspective, the observation of lepton number violating processes would signal the presence of physics beyond the SM, complementing their implications for neutrino masses and the Majorana character of neutrinos. In particular, the observation of $\Delta L = 2$ interactions at a high-energy collider such as the $\mu$TRISTAN would directly point to new degrees of freedom at the TeV scale, instead of decoupled, ultra-heavy states typical of vanilla seesaw realizations. Such low-scale manifestations of LNV can in turn indicate towards non-minimal extensions where neutrino mass is either generated radiatively, or through flavor-structured dynamics that suppress the Weinberg operator, allowing sizable higher-dimensional effects. More broadly, LNV is also a key ingredient in addressing fundamental open questions most notably the origin of the matter-antimatter asymmetry of the Universe via leptogenesis \cite{Fukugita:1986hr}, subsequently converted into a baryon asymmetry. As emphasized in \cite{Deppisch:2013jxa}, a non-zero LNV cross section would imply efficient lepton number washout near the electroweak phase transition, thus effectively falsifying high-scale leptogenesis scenarios by violating the out-of-equilibrium condition, while the absence of LNV signal leaves leptogenesis unconstrained. Consequently, establishing LNV at colliders would not only reveal the presence of accessible NP, but also provide critical clues about the underlying structure of lepton number breaking and its possible role in shaping the observed properties of the Universe.}

The paper is organized as follows: Sec. \ref{sec:effective} introduces the effective operators relevant to our analysis. In Sec. \ref{sec:constraints}, we review existing constraints on these operators. Secs. \ref{sec:collider} and \ref{sec:sensitivity} are devoted to the collider phenomenology and the projected sensitivity of future experiments. {In Sec. \ref{sec:fcccomp}, we compare our projected sensitivities with those expected at the FCC-hh, while Sec. \ref{sec:nptoeft} presents an explicit NP scenario that generates one of the effective operators and examines its phenomenological implications.} Finally, we summarize and conclude in Sec. \ref{sec:conclusion}.

\section{LNV operators in SMEFT}
\label{sec:effective}

We focus on $\Delta L = 2$ dimension 7 operators. Among 12 such operators, eight contribute at the tree level to the processes $W^{+}W^{+}$ and $W^{+}qq'$ at the 
$\mu$TRISTAN. These operators, along with their relevant interaction vertices, are listed in Tab. \ref{tab:d7ops}. The operator $\mathcal{O}_{LH}: \epsilon_{ij} \epsilon_{mn} (\overline{L^{c}_{i}} L_{m}) H_{j} H_{n} (H^{\dagger} H)$ also contributes to $W^{+}W^{+}/\;W^{+}qq'$ production (via $\nu \nu$ vertex), but it generates the same vertices as the Weinberg operator unless one considers interactions involving more than two Higgs fields and is therefore redundant for the purposes of our analysis.

\begin{table*}[htb!]
    \centering
    \renewcommand{\arraystretch}{1.2}{
    \begin{tabular}{|>{\centering\arraybackslash}p{2cm}|>{\centering\arraybackslash}p{2cm}|>{\centering\arraybackslash}p{6cm}|>{\centering\arraybackslash}p{4cm}|}
        \hline
         \multicolumn{3}{|c|}{Operators} & Vertices \\
        \hline
        \multirow{2}*{$\Psi^{2} H^{2} D^{2}$} & $\mathcal{O}_{LHD1}$ & $\epsilon_{ij} \epsilon_{mn} (\overline{L^{c}_{i}} D^{\mu} L_{j}) (H_{m} D_{\mu} H_{n})$ & $\mu \nu_{\mu} W,\; \mu \mu WW$ \\
        & $\mathcal{O}_{LHD2}$ & $\epsilon_{im} \epsilon_{jn} (\overline{L^{c}_{i}} D^{\mu} L_{j}) (H_{m} D_{\mu} H_{n})$ & $\mu\nu_{\mu} W$ \\ \hline
        $\Psi^{2} H^{3} D$ & $\mathcal{O}_{LHDe}$ & $\epsilon_{ij} \epsilon_{mn} (\overline{L^{c}_{i}} \gamma^{\mu} e) H_{j} (H_{m} i D_{\mu} H_{n})$ & $\mu \nu_{\mu} W$ \\ \hline
        $\Psi^{4} D$ & $\mathcal{O}_{LLduD}$ & $\epsilon_{ij} (\overline{d} \gamma^{\mu} u)(\overline{L^{c}_{i}} i D_{\mu} L_{j}) $ & $\mu\nu_{\mu}qq', \; \mu \mu Wqq'$ \\ \hline
        \multirow{4}*{$\Psi^{4} H$} & $\mathcal{O}_{LLQdH1}$ & $\epsilon_{ij} \epsilon_{mn} (\overline{d} L_{i})(\overline{Q^{c}_{j}} L_{m})H_{n}$ & $\mu\nu_{\mu} qq'$\\
        & $\mathcal{O}_{LLQdH2}$ & $\epsilon_{im} \epsilon_{jn} (\overline{d} L_{i})(\overline{Q^{c}_{j}} L_{m})H_{n}$ & $\mu \nu_{\mu} qq'$ \\
        & $\mathcal{O}_{LLQuH}$ & $\epsilon_{ij} (\overline{Q_{m}} u)(\overline{L^{c}_{m}} L_{i})H_{j}$ & $\mu \nu_{\mu} qq'$ \\
        & $\mathcal{O}_{LeudH}$ & $\epsilon_{ij} (\overline{L^{c}_{i}}\gamma^{\mu} e)(\overline{d} \gamma_{\mu} u)H_{j}$ & $\mu \nu_{\mu} qq'$ \\
        \hline
    \end{tabular}}
    \caption{$\Delta L = 2$ dimension 7 SMEFT operators contributing to $W^{+}W^{+}(jj)\;/\;W^{+}jj$ signal process at the $\mu$TRISTAN and the corresponding effective vertices. The corresponding WCs have the same subscripts as the effective operators.}
    \label{tab:d7ops}
\end{table*}

It is important to note that the operators listed in Tab. \ref{tab:d7ops} carry flavor indices. In our analysis, we focus specifically on muon-flavored operators, i.e., those with flavor structures $\mu \mu$ and $\mu \mu qq'$, with no lepton flavor violation. While quark flavor mixing is allowed, we restrict the quark indices to the first two generations. The inclusion of third-generation quarks can lead to additional signal channels, which we leave for future investigation. The operators are further categorized into four classes based on their Lorentz structure: $\Psi^2 H^2 D^2$, $\Psi^2 H^3 D$, $\Psi^4 D$, and $\Psi^4 H$. The latter two classes, involving four fermion fields, are commonly referred to as four-fermion operators.

\section{Constraints on LNV operators}
\label{sec:constraints}
Before exploring collider signatures at the $\mu$TRISTAN, it is essential to consider existing constraints on the lepton number violating operators. These constraints arise from a variety of sources, including low-energy processes, precision measurements, and previous collider studies. A comprehensive analysis of existing constraints on LNV operators is provided in \cite{Fridell:2023rtr}. The most stringent bounds arise from the non-observation of neutrinoless double beta decay, but these are specific to operators involving the electron flavor and do not directly constrain muon-flavored operators unless a flavor symmetry is assumed. Additional constraints on multi-flavor operators come from processes such as coherent elastic neutrino-nucleus scattering \cite{PhysRevD.9.1389,Kerman:2016jqp,Lindner:2016wff,Bolton:2021pey,Bischer:2019ttk,Bolton:2020xsm,Li:2020lba,AristizabalSierra:2018eqm}, neutrino oscillations \cite{Bolton:2019wta,Danko:2009qw,Formaggio:2012cpf,Bahcall:1996qv}, rare meson and heavy lepton decays \cite{Liao:2019gex,Liao:2020roy,Zhou:2021lnl,Atre:2005eb,Abada:2019bac}, and measurements of the neutrino magnetic moment \cite{Bell:2006wi,Giunti:2014ixa,Canas:2015yoa,AristizabalSierra:2021fuc,deGouvea:2022znk}. However, for operators involving only muon flavors, the bounds from these low-energy observables remain relatively weak.

Constraints on these LNV operators from the LHC have been explored primarily through phenomenological studies \cite{Fridell:2023rtr,Cepedello:2017lyo,delAguila:2012nu,Fuks:2020att,Aoki:2020til}, focusing on signatures like same-sign dilepton plus jets, etc. A detailed phenomenological study in \cite{Fridell:2023rtr} recasts the ATLAS analysis \cite{ATLAS:2023cjo} to investigate same-sign muon signal processes of the form $pp \rightarrow \mu^{\pm} \mu^{\pm} jj$, where $j$ stands for light quark jets. The resulting bounds on the relevant LNV operator coefficients (in TeV$^{-3}$) and the corresponding effective scale $\Lambda$ (in TeV) are summarized in Tab. \ref{tab:lhc_wc}.
\begin{table}[htb!]
    \centering
    \renewcommand{\arraystretch}{1.2}{
    \begin{tabular}{|>{\centering\arraybackslash}p{2cm}|>{\centering\arraybackslash}p{2cm}|>{\centering\arraybackslash}p{3cm}|}
        \hline
        & \multicolumn{2}{c|}{LHC 13 TeV 139 fb$^{-1}$} \\ \cline{2-3} 
        \multirow{-2}*{Operators} & $\Lambda$ (TeV) & $C_i/\Lambda^3$ (TeV$^{-3}$) \\
        \hline
        $\mathcal{O}_{LHD1}$ & 1.100 & $7.513 \times 10^{-1}$ \\
        $\mathcal{O}_{LHD2}$ & 0.075 & $2.370 \times 10^{3}$  \\
        $\mathcal{O}_{LHDe}$ & 0.210 & $1.080 \times 10^{2}$  \\
        $\mathcal{O}_{LLduD}$ & 5.000 & $8.000 \times 10^{-3}$ \\
        $\mathcal{O}_{LLQdH1}$ & 1.100 & $7.513 \times 10^{-1}$ \\
        $\mathcal{O}_{LLQdH2}$ & 0.900 & $1.372 \times 10^{0}$ \\
        $\mathcal{O}_{LLQuH}$ & 1.400 & $3.644 \times 10^{-1}$ \\
        $\mathcal{O}_{LeudH}$ & 1.100 & $7.513 \times 10^{-1}$ \\
        \hline
    \end{tabular}}
    \caption{Upper limits on $\Lambda$ and WCs for different dimension 7 SMEFT operators from LHC 13 TeV 139 fb$^{-1}$ \cite{Fridell:2023rtr,ATLAS:2023cjo}.}
    \label{tab:lhc_wc}
\end{table}

In the LHC context, four-fermion operators tend to receive stronger bounds, primarily due to the presence of quark fields enabling direct production processes. Among these, the operator $\mathcal{O}_{LLduD}$ is the most tightly constrained, as it contributes through a contact interaction, enhancing the signal rate at the tree level. Operators belonging to the $\Psi^2 H^2 D^2$ and $\Psi^2 H^3 D$ classes predominantly contribute to the signal process through $W$ boson fusion (WBF). Among them, the operator $\mathcal{O}_{LHD1}$ is more strongly constrained, as its contribution involves fewer propagators in the production diagram, leading to a less suppressed signal rate. While the LHC offers a high-energy environment for such probes, its sensitivity to the $\Delta L = 2$ dimension-7 operators is limited by backgrounds and relatively small cross sections. As a result, the bounds derived from these studies are relatively weak and leave ample room for exploration at future same-sign lepton colliders.
\section{\boldmath $W^{+}W^{+}/\;W^{+}qq'$ production at $\mu$TRISTAN}
\label{sec:collider}

The Feynman diagrams for the signal processes $W^{+}W^{+}/\;W^{+}qq'$ arising from the dimension seven SMEFT operators are shown in Fig. \ref{fig:lnv-wjj}. The operators $\mathcal{O}_{LHD1}$ contribute to Fig. \ref{fig:fda} and \ref{fig:fdc}, and $\mathcal{O}_{LLduD}$ contribute to Fig. \ref{fig:fdb} and \ref{fig:fdd}. The operators $\mathcal{O}_{LHD2}$ and $\mathcal{O}_{LHDe}$ contribute to the diagram in Fig. \ref{fig:fdc}, while the four-fermion operators of class $\Psi^4 H$ contribute through Fig. \ref{fig:fdd}.
\begin{figure}[htb!]
\centering
\subfloat[]{\begin{tikzpicture}[baseline={(current bounding box.center)},style={scale=0.75, transform shape}]
\begin{feynman}
\vertex(a);
\vertex[above left =1.25cm and 1.25cm of a] (a1){$\mu^{+}$};
\vertex[below left =1.25cm and 1.25cm of a] (a2){$\mu^{+}$};
\vertex[above right =1.25cm and 1.25cm of a] (b1){$W^{+}$};
\vertex[below right =1.25cm and 1.25cm of a] (b2){$W^{+}$};
\diagram*{
(a) -- [ fermion, arrow size=0.7pt, thick] (a1),
(a) -- [ fermion, arrow size=0.7pt, thick] (a2),
(a) -- [ boson, arrow size=0.7pt, thick] (b1),
(a) -- [ boson, arrow size=0.7pt, thick] (b2)};
\end{feynman}
\node at (a)[black,fill,style=black,inner sep=3pt]{};
\end{tikzpicture}
\label{fig:fda}}
\subfloat[]{\begin{tikzpicture}[baseline={(current bounding box.center)},style={scale=0.75, transform shape}]
\begin{feynman}
\vertex(a);
\vertex[above left =1.25cm and 1.25cm of a] (a1){$\mu^{+}$};
\vertex[below left =1.25cm and 1.25cm of a] (a2){$\mu^{+}$};
\vertex[above right =1.25cm and 1.25cm of a] (b1){$W^{+}$};
\vertex[right =1.5cm of a] (b2){$q$};
\vertex[below right =1.5cm and 1.5cm of a] (b3){$q'$};
\diagram*{
(a) -- [ fermion, arrow size=0.7pt, thick] (a1),
(a) -- [ fermion, arrow size=0.7pt, thick] (a2),
(a) -- [ boson, arrow size=0.7pt, thick] (b1),
(b2) -- [ fermion, arrow size=0.7pt, thick] (a),
(a) -- [ fermion, arrow size=0.7pt, thick] (b3)};
\end{feynman}
\node at (a)[black,fill,style=black,inner sep=3pt]{};
\end{tikzpicture}
\label{fig:fdb}} \\
\subfloat[]{\begin{tikzpicture}[baseline={(current bounding box.center)},style={scale=0.75, transform shape}]
\begin{feynman}
\vertex(a);
\vertex[left =1.5cm of a] (a1){$\mu^{+}$};
\vertex[right =1.5cm of a] (a2){$W^{+}$};
\vertex[below = 3.3cm of a] (c);
\vertex[left =1.5cm of c] (c1){$\mu^{+}$};
\vertex[right =1.5cm of c] (c2){$W^{+}$};
\diagram*{
(a) -- [ fermion, arrow size=0.7pt, thick] (a1),
(a2) -- [ boson, arrow size=0.7pt, thick] (a),
(c) -- [ fermion, arrow size=0.7pt, thick] (c1),
(c2) -- [ boson, arrow size=0.7pt, thick] (c),
(c) -- [ fermion, arrow size=0.7pt, edge label = $\overline{\nu}$, thick] (a)
};
\end{feynman}
\node at (c)[black,fill,style=black,inner sep=3pt]{};
\end{tikzpicture}
\label{fig:fdc}}
\subfloat[]{\begin{tikzpicture}[baseline={(current bounding box.center)},style={scale=0.75, transform shape}]
\begin{feynman}
\vertex(a);
\vertex[left =1.5cm of a] (a1){$\mu^{+}$};
\vertex[right =1.5cm of a] (a2){$W^{+}$};
\vertex[below = 3.3cm of a] (c);
\vertex[left =1.5cm of c] (c1){$\mu^{+}$};
\vertex[above right =1.5cm and 1.5cm of c] (c3){$q$};
\vertex[right =1.5cm of c] (c2){$q'$};
\diagram*{
(a) -- [ fermion, arrow size=0.7pt, thick] (a1),
(a2) -- [ boson, arrow size=0.7pt, thick] (a),
(c) -- [ fermion, arrow size=0.7pt, thick] (c1),
(c) -- [ fermion, arrow size=0.7pt, thick] (c2),
(c3) -- [ fermion, arrow size=0.7pt, thick] (c),
(c) -- [ fermion, arrow size=0.7pt, edge label = $\overline{\nu}$, thick] (a)
};
\end{feynman}
\node at (c)[black,fill,style=black,inner sep=3pt]{};
\end{tikzpicture}
\label{fig:fdd}}
\caption{Feynman graphs corresponding to $W^{+}W^{+}/\, W^{+}qq'$ signal process at the $\mu$TRISTAN. Black square dot refers to the effective vertices arising from the 
dimension seven SMEFT operators.}
\label{fig:lnv-wjj}
\end{figure}

\begin{figure*}[htb!]
    \centering
    \includegraphics[width=\linewidth]{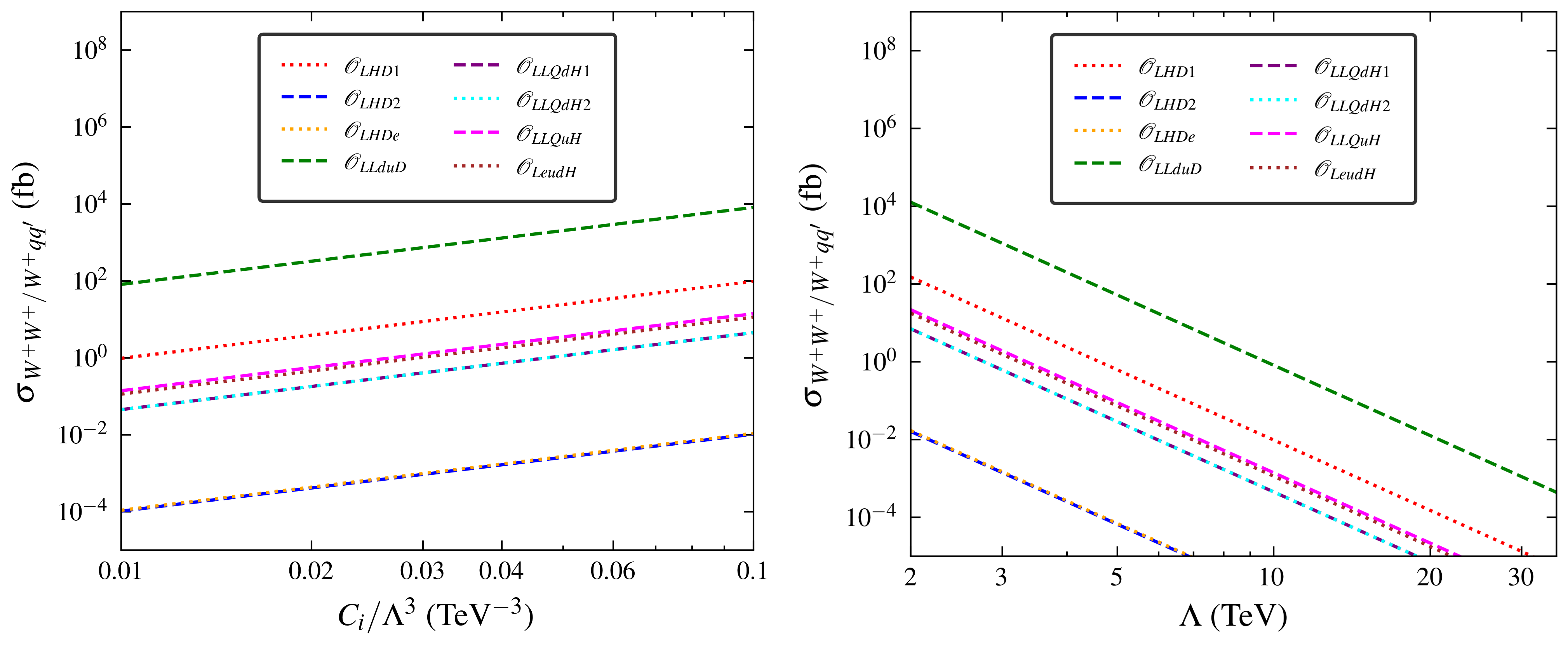}
    \caption{Production cross section of $\mu^+ \mu^+ \rightarrow W^{+}W^{+}/\;W^{+}qq'$ process as a function of the WCs, $C_i/\Lambda^{3}$ (\textit{left}), and EFT scale ($\Lambda$) with $C_i$ set to 1 (\textit{right}), at the $\mu$TRISTAN $\sqrt{s} = 2$ TeV.}
    \label{fig:xs}
\end{figure*}

The production cross sections for the signal processes at $\mu$TRISTAN with $\sqrt{s} = 2$ TeV are shown in Fig. \ref{fig:xs} with respect to the NP scale variation; 
$C_i/\Lambda^{3}$ (\textit{left}) and $\Lambda$ with $C_i=1$ (\textit{right}). Among the operators, $\mathcal{O}_{LHD1}$ and $\mathcal{O}_{LLduD}$ yield the largest cross sections, primarily due to their dominant contributions arising from contact interaction diagrams (Fig. \ref{fig:fda} and \ref{fig:fdb}, respectively), which scale with the center-of-mass energy. Between the two, $\mathcal{O}_{LLduD}$ results in a higher cross section due to the presence of lighter final states. While the four-fermion operators of class $\Psi^4 H$ also feature light final states, their cross sections are suppressed by propagator effects. In contrast, the operators $\mathcal{O}_{LHD2}$ and $\mathcal{O}_{LHDe}$ are further suppressed due to both propagator dependence and heavier final states.

\subsection{Collider analysis}
For our analysis of the signal processes $W^{+}W^{+}/\;W^{+}qq'$, we consider the final-state signature with 2 fat jets ($2J$). In the $W^{+}W^{+}$ process, fat jets originate from the hadronic decays of the $W$ bosons. For the $W^{+}qq'$ process, the final state typically contains two or three fat jets, one from the hadronic decay of the $W$ boson and one or two additional fat jets from the final-state quarks, depending on their degree of collimation. To maintain consistency across both processes, we restrict the combined analysis to events with exactly two fat jets. The dominant Standard Model background arises from the process $W^{+}W^{+}\overline{\nu}\,\overline{\nu}$. Subdominant backgrounds include channels such as $\mu^+ \overline{\nu} W^{+}$, $\mu^+ \mu^+ Z$, and $W^{+} Z\,\mu^+ \overline{\nu}$, where one or more muons escape detection. However, these subdominant contributions are strongly suppressed and are completely removed by the cutflow strategy described later.

The SMEFT contribution is estimated via implementing the operators using \texttt{FeynRules} \cite{Alloul:2013bka}. Four-fermion flow-violating vertices are handled in a manner similar to the approach used in \cite{Dong:2011rh}. Signal and dominant SM background events are generated with \texttt{MG5\_aMC} \cite{Alwall:2014hca}, followed by parton showering using \texttt{Pythia8} \cite{Sjostrand:2014zea}. Detector-level simulation is performed with \texttt{Delphes3} \cite{deFavereau:2013fsa}, utilizing the default \texttt{Delphes} detector card provided in \texttt{MG5\_aMC}. The basic object selection criteria based on transverse momentum ($p_T$) and pseudorapidity ($\eta$) are as follows:
\textbf{Electrons:} $p_{T} > 10$ GeV and $|\eta| < 2.5$, with identification efficiency: 0.95 for $|\eta| < 1.5$ and 0.85 for $1.5 < |\eta| < 2.5$.
\textbf{Muons:} $p_{T} > 10$ GeV and $|\eta| < 2.4$, with identification efficiency: 0.95.
\textbf{Photons:} $p_{T} > 10$ GeV and $|\eta| < 2.5$, with identification efficiency: 0.95 for $|\eta| < 1.5$ and 0.85 for $1.5 < |\eta| < 2.5$.
\textbf{Fat jets:} The fat jets are clustered using \texttt{FastJet3} \cite{Cacciari:2011ma} with the anti-$k_t$ algorithm \cite{Cacciari:2008gp} within the \texttt{Delphes3} framework. The minimum $p_{T}$ required to form a fat jet is set to 200 GeV, and the jet radius is chosen to be 0.8. The parameters for Pruning \cite{Ellis:2009me}, Trimming \cite{Krohn:2009th}, and SoftDrop \cite{Larkoski:2014wba} are set to the default values provided by \texttt{Delphes3}. For event selection, we require exactly two fat jets ($N_J = 2$) and no electrons, muons, or photons ($N_e = 0$, $N_\mu = 0$, $N_\gamma = 0$).

\begin{figure*}[htb!]
    \centering
    \includegraphics[width=\linewidth]{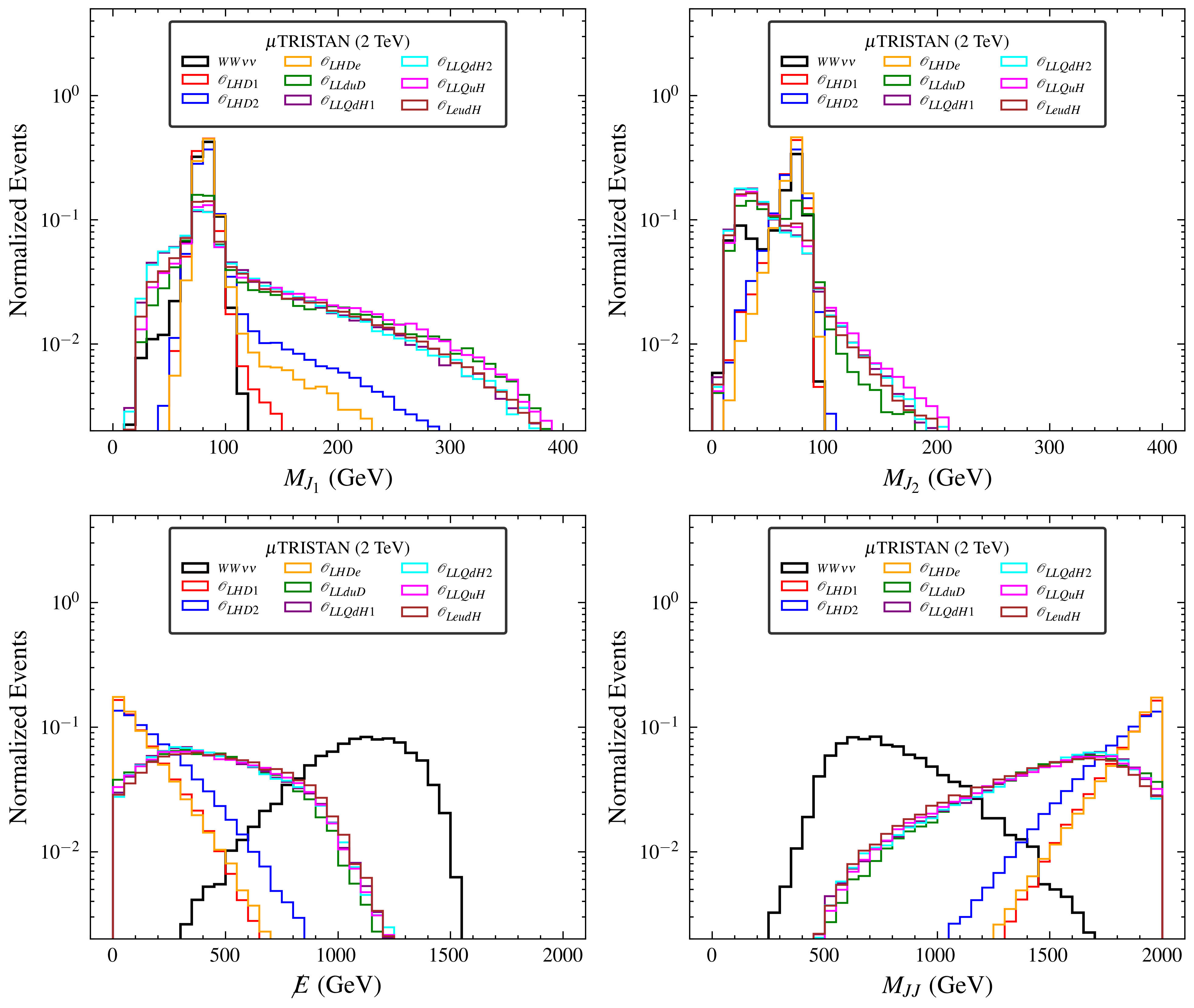}
    \caption{Kinematic distributions corresponding to EFT signals (colored) and the dominant SM background, $W^{+}W^{+}\overline{\nu}\,\overline{\nu}$ (black) at the $\mu$TRISTAN 2 TeV.}
    \label{fig:dist}
\end{figure*}

The relevant kinematic distributions for the signal and the background are shown in Fig. \ref{fig:dist}. The definition of the variables are as follows:
\begin{itemize}
    \item $M_{J_1}$ and $M_{J_2}$ are the masses of the heavier and the lighter fat jets, respectively, defined as:
    \begin{equation}
        M_{J} = \sqrt{\left(\sum_{i} p_{i}\right)^2}\,,
    \end{equation}
    where, $p_{i}$ are the 4-momenta of the reconstructed constituents of the fat jet.
    \item The missing energy, $\slashed{E}$ is defined as
    \begin{equation}
        \slashed{E} = \sqrt{s} - \sum_{i} E_{i},
    \end{equation}
    where, $E_{i}$ are the energies of the visible final state particles.
    \item Invariant mass of the fat jet pair, $M_{JJ}$ is defined as
    \begin{equation}
        M_{jj} = \sqrt{\left(p_{J_1} + p_{J_2}\right)^{2}},
    \end{equation}
    where, $p_{J_1}$ and $p_{J_2}$ are 4-momenta of the heavier and lighter fat jet, respectively.
\end{itemize}
To enhance the signal-to-background separation, we apply the following sequential kinematic cuts, in addition to the basic detector-level and signal selection cuts:
\begin{itemize}
    \itemsep-0em
    \item $\mathcal{C}_{1}:\; M_{J_2} > 10\; {\rm GeV},$
    \item $\mathcal{C}_{2}:\; \slashed{E} < 400\; {\rm GeV},$
    \item $\mathcal{C}_{3}:\; M_{JJ} > 1600\; {\rm GeV}.$
\end{itemize}
The mass of the lighter fat jet is required to be greater than 10 GeV to suppress backgrounds involving fat jets originating from final-state radiation or secondary hadronization. This cut completely eliminates background contributions from the $\mu^+ \overline{\nu} W^+$ and $\mu^+ \mu^+ Z$ channels, which survive the basic detector-level and signal selection cuts. Notably, applying $M_{J_2} > 10$ GeV automatically guarantees that $M_{J_1}$ also exceeds 10 GeV, since jets are ordered by decreasing mass. The most effective discriminating variable turns out to be the missing energy, which tends to be larger for SM backgrounds due to multiple invisible neutrinos in the final state. The $W^{+}W^{+}$ signal process offers better separation from the background compared to the $W^{+}qq'$ channel. Similarly, the invariant mass of the fat jet pair ($M_{JJ}$) peaks near the CM energy for signal events, whereas for background processes, the distribution shifts to lower values due to energy carried away by neutrinos. The cutflow for both signal and background processes, corresponding to individual operator scenarios, is presented in Tab. \ref{tab:ceff}. {The WCs are set to $C_i/\Lambda^3 = 0.1$ TeV$^{-3}$ for all operators, corresponding to an effective scale of $\Lambda \sim 2.15$ TeV (assuming $C_i \sim 1$). This choice satisfies the EFT validity condition (detailed in Appendix \ref{sec:appX}) for collider studies, which requires $\Lambda > \sqrt{s}$; in our case, $\sqrt{s} = 2$ TeV. It should be noted that for the operator $\mathcal{O}_{LLduD}$, the stringent LHC constraints (see Tab.~\ref{tab:lhc_wc}) require a smaller value of $C_{LLduD}/\Lambda^{3}$. Nevertheless, we set $C_{LLduD}/\Lambda^3 = 0.1~\text{TeV}^{-3}$ to maintain consistency with the values used for the other operators.} The final cut efficiency, $\epsilon$, is defined as the ratio of the cross section after the final cut ($\mathcal{C}_3$), $\sigma_{3}$, to the production cross section, $\sigma_{\tt prod}$. Operators contributing to the $W^+W^+$ process exhibit higher cut efficiencies compared to those contributing to the $W^+qq'$ process, owing to their better separation from the background.
\begin{table*}[htb!]
    \centering
    \renewcommand{\arraystretch}{1.2}{
    \begin{tabular}{|>{\centering\arraybackslash}p{3cm}|>{\centering\arraybackslash}p{2.75cm}|>{\centering\arraybackslash}p{2.75cm}|>{\centering\arraybackslash}p{2.75cm}|>{\centering\arraybackslash}p{2.75cm}|}
    \hline
        \multirow{2}*{Operators} & \multicolumn{3}{c|}{Cross sections (fb)} & \multirow{2}*{Cut efficiency, $\epsilon$} \\ \cline{2-4}
         & $\sigma_{1}$ & $\sigma_{2}$ & $\sigma_{3}$ & \\ \hline
        $\mathcal{O}_{LHD1}$ & $5.091 \times 10^{1}$ & $4.869 \times 10^{1}$ & $4.812 \times 10^{1}$ & 0.7390  \\
        $\mathcal{O}_{LHD2}$ & $6.244 \times 10^{-3}$ & $5.495 \times 10^{-3}$ & $5.378 \times 10^{-3}$ & 0.7814 \\
        $\mathcal{O}_{LHDe}$ & $4.499 \times 10^{-3}$ & $4.280 \times 10^{-3}$ & $4.229 \times 10^{-3}$ & 0.5777 \\
        $\mathcal{O}_{LLduD}$ & $2.089 \times 10^{3}$ & $1.012 \times 10^{3}$ & $9.405 \times 10^{2}$ & 0.1728 \\
        $\mathcal{O}_{LLQdH1}$ & $1.310 \times 10^{0}$ & $6.204 \times 10^{-1}$ & $5.787 \times 10^{-1}$ & 0.1921 \\
        $\mathcal{O}_{LLQdH2}$ & $1.304 \times 10^{0}$ & $6.143 \times 10^{-1}$ & $5.714 \times 10^{-1}$ & 0.1898 \\
        $\mathcal{O}_{LLQuH}$ & $3.564 \times 10^{0}$ & $1.659 \times 10^{0}$ & $1.541 \times 10^{0}$ & 0.1645 \\
        $\mathcal{O}_{LeudH}$ & $3.172 \times 10^{0}$ & $1.363 \times 10^{0}$ & $1.264 \times 10^{0}$ & 0.1644 \\
        \hline
        SM Background & $3.047 \times 10^{0}$ & $2.714 \times 10^{-2}$ & $2.255 \times 10^{-2}$ & 0.0006 \\ \hline
    \end{tabular}}
    \caption{Cross sections ($\sigma_i$) after each sequential cut $\mathcal{C}_i$, along with the final cut efficiency ($\epsilon = \sigma_3 / \sigma_{\tt prod}$) for the SM background processes and individual LNV operators contributing to the $W^{+}W^{+}/\;W^{+}qq'$ signal at the $\mu$TRISTAN with $\sqrt{s} = 2$ TeV with $\mathfrak{L}_{\rm int} = 1$ ab$^{-1}$. The WCs are fixed at $C_i/\Lambda^3 = 0.1$ TeV$^{-3}$ for all operators.}
    \label{tab:ceff}
\end{table*}
\section{Projected sensitivities of operators}
\label{sec:sensitivity}
To assess the discovery potential of these operators at $\mu$TRISTAN, we perform a binned $\chi^2$ analysis on key observables to estimate the projected sensitivities of the LNV operator coefficients. This allows us to determine the exclusion limits at various confidence levels, assuming an integrated luminosity of 1 ab$^{-1}$. Different formulations of the $\chi^2$ test exist in the literature, each becoming equivalent under certain limits. For our analysis, we adopt the Neyman definition \cite{Ji:2019yca}:
\begin{equation}
    \chi^{2} = \sum^{\rm bins}_{r} \left( \frac{\mathcal{Q}_{r} (C_{i}) - \mathcal{Q}_{r} (0)}{\sqrt{\mathcal{Q}_{r} (0)}} \right)^{2},
\end{equation}
where $\mathcal{Q}_{r} (C_{i})$ and $\mathcal{Q}_{r} (0)$ denote the number of events in the $r^{\rm th}$ bin of a chosen observable, with and without the presence of LNV operators, respectively. In our study, we use the differential observable:
\begin{equation}
    \mathcal{Q}_{r} (C_{i}) = \int_{r} \mathfrak{L}_{\rm int} \times \left(\frac{d\sigma (C_{i})}{d\cos{\theta_1}} \right)\; d (\cos{\theta_1}),
\end{equation}
where $\theta_1$ is the angle between the heavier jet and the beam axis. The integration is performed over the range corresponding to the width of the $r^{\rm th}$ bin. For our analysis, the $\cos{\theta_1}$ distribution is divided into five equally spaced bins, i.e., $r=1,2,\ldots,5$. Note that the $\chi^2$ definition used here accounts only for statistical uncertainties and does not include additional systematic uncertainties that could affect the estimations. In experimental analyses, such uncertainties are typically handled using nuisance parameters \cite{ParticleDataGroup:2024cfk,Xia:2018cpd}, but implementing them requires detailed technical inputs beyond the scope of this study and is therefore not considered.
\subsection{One operator analysis}
\begin{figure*}[htb!]
    \centering
    \includegraphics[width=\linewidth]{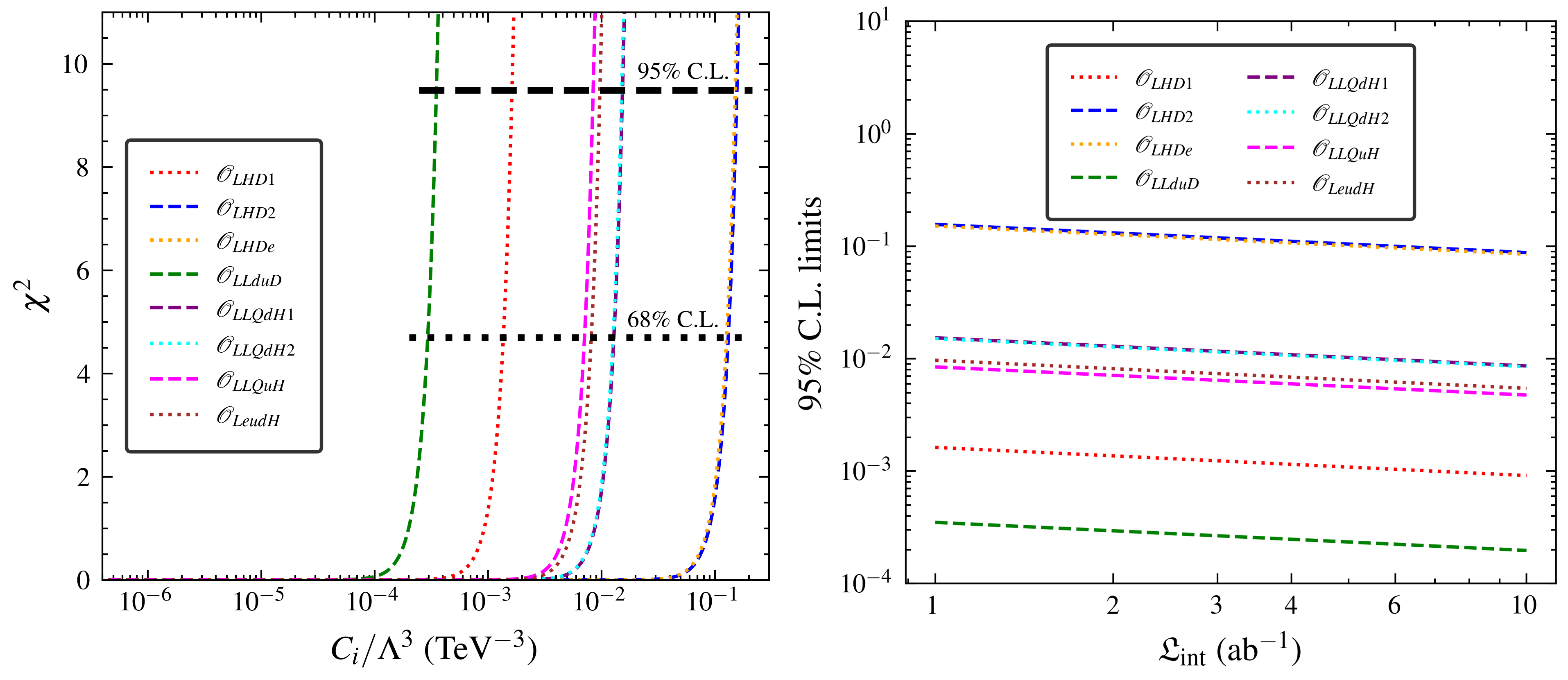}
    \caption{\textit{Left:} $\chi^{2}$ sensitivity curves for the LNV operator coefficients corresponding to $W^+W^+/\;W^+qq'$ production at the $\mu$TRISTAN with $\sqrt{s} = 2$ TeV with $\mathfrak{L}_{\rm int} = 1$ ab$^{-1}$. The black dashed (dotted) horizontal line indicates the 95\% (68\%) C.L. exclusion limit. \textit{Right:} 95\% C.L. limits on operators as a function of integrated luminosity, $\mathfrak{L}_{\rm int}$.}
    \label{fig:1d-oot}
\end{figure*}

We begin by analyzing individual operator scenarios. For an $m$-parameter fit, the degrees of freedom (d.o.f.) in a $\chi^2$ analysis using a non-normalized observable is given by $n - m$, where $n$ is the number of bins \cite{ParticleDataGroup:2024cfk}. In the case of a single parameter fit with 5 bins, the d.o.f. is 4. The corresponding $\chi^2$ values for 68\% C.L. and 95\% C.L. are 4.695 and 9.488, respectively. The $\chi^2$ sensitivity curves for various LNV operators are presented in Fig. \ref{fig:1d-oot} (\textit{left}), with the 95\% (68\%) confidence level exclusion limits indicated by black dashed (dotted) lines. The exclusion limits on $C_{i}/\Lambda^{3}$ for each operator are listed in Tab. \ref{tab:oot1d}. Consistent with the LHC constraints, the strongest bound is obtained for the operator $\mathcal{O}_{LLduD}$. Notably, there is a substantial improvement in the bounds for operators contributing to the $W^+W^+$ channel compared to existing LHC limits, with similarly enhanced sensitivity for the four-fermion operators contributing to the $W^+qq'$ process. Fig. \ref{fig:1d-oot} (\textit{right}) illustrates the improvement of exclusion limits with increasing luminosity. While the gains are substantial, we restrict our analysis to 1 ab$^{-1}$ to remain consistent with the projected target for $\mu$TRISTAN \cite{Hamada:2022mua}.

\begin{table}[htb!]
    \centering
    \renewcommand{\arraystretch}{1.2}{
    \begin{tabular}{|>{\centering\arraybackslash}p{2cm}|>{\centering\arraybackslash}p{2.5cm}|>{\centering\arraybackslash}p{2.5cm}|}
        \hline
        \multirow{2}*{Operators} & \multicolumn{2}{c|}{Bounds on $C_{i}/\Lambda^{3}$ (TeV$^{-3}$)} \\ \cline{2-3}
         & $68\%$ C.L. & $95\%$ C.L. \\ \hline
        $\mathcal{O}_{LHD1}$ & $1.364 \times 10^{-3}$ & $1.627 \times 10^{-3}$ \\
        $\mathcal{O}_{LHD2}$ & $1.311 \times 10^{-1}$ & $1.563 \times 10^{-1}$ \\
        $\mathcal{O}_{LHDe}$ & $1.267 \times 10^{-1}$ & $1.511 \times 10^{-1}$ \\
        $\mathcal{O}_{LLduD}$ & $2.941 \times 10^{-4}$ & $3.507 \times 10^{-4}$ \\
        $\mathcal{O}_{LLQdH1}$ & $1.286 \times 10^{-2}$ & $1.534 \times 10^{-2}$ \\
        $\mathcal{O}_{LLQdH2}$ & $1.266 \times 10^{-2}$ & $1.510 \times 10^{-2}$ \\
        $\mathcal{O}_{LLQuH}$ & $7.088 \times 10^{-3}$ & $8.451 \times 10^{-3}$ \\
        $\mathcal{O}_{LeudH}$ & $8.130 \times 10^{-3}$ & $9.693 \times 10^{-3}$ \\
         \hline
    \end{tabular}}
    \caption{68\% and 95\% C.L. upper limits on $C_{i}/\Lambda^{3}$ (TeV$^{-3}$) corresponding to $W^+W^+/\;W^+qq'$ production at the $\mu$TRISTAN with $\sqrt{s}=$2 TeV, and $\mathfrak{L}_{\rm int} = 1$ ab$^{-1}$.}
    \label{tab:oot1d}
\end{table}
\subsection{Two operator analysis}
\begin{figure*}[htb!]
    \centering
    \includegraphics[width=\linewidth]{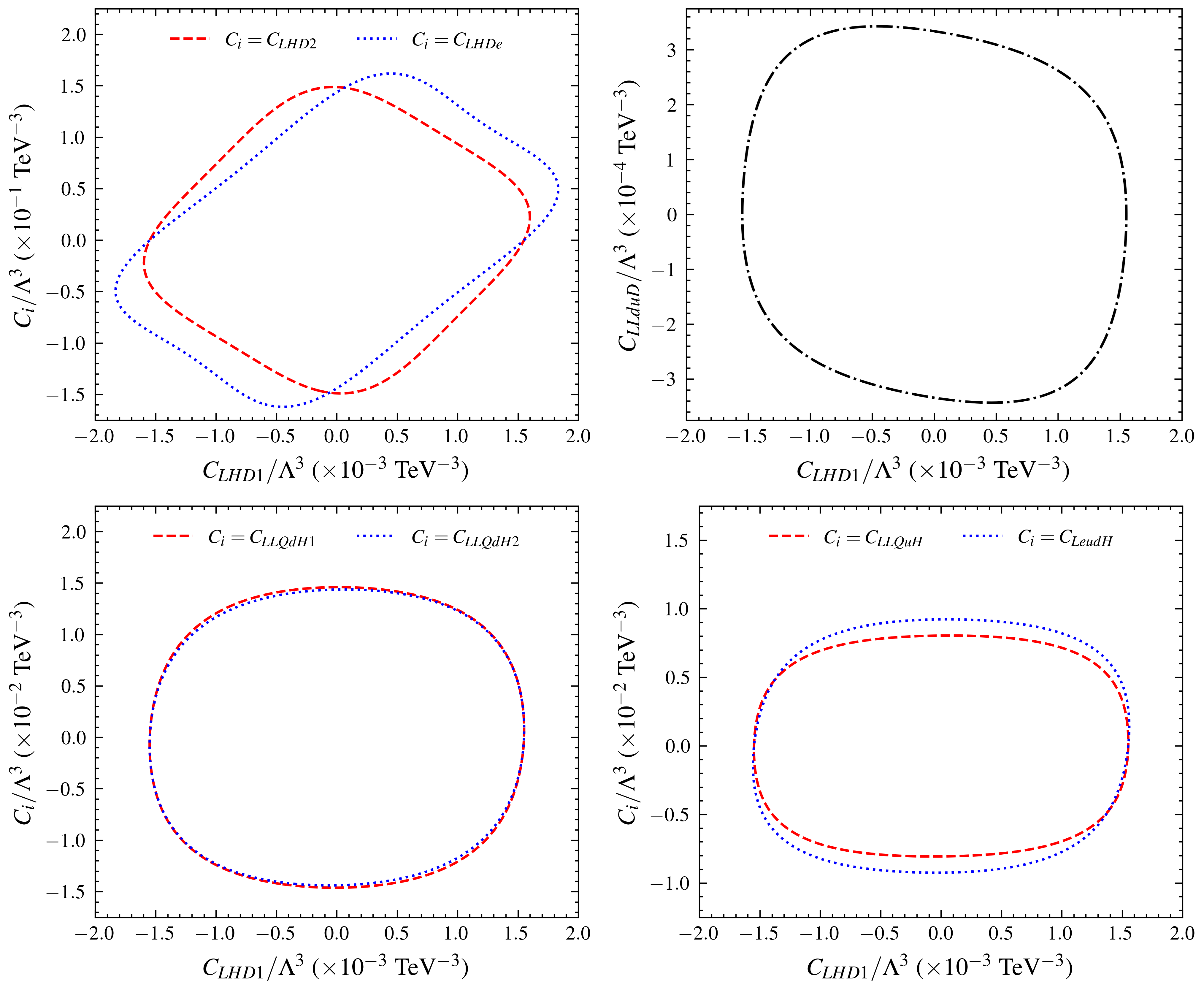}
    \caption{Correlated 95\% C.L. $\chi^{2}$ sensitivity contours for the LNV operator $\mathcal{O}_{LHD1}$ in combination with other operators, derived from $W^+W^+$ and $W^+qq'$ production at the $\mu$TRISTAN with $\sqrt{s} = 2$ TeV and an integrated luminosity of $\mathfrak{L}_{\rm int} = 1$ ab$^{-1}$.}
    \label{fig:2d-oot}
\end{figure*}

Following the approach of the individual operator analysis, we now extend the study to two-operator scenarios. For a two-parameter fit, the d.o.f. is 3, with the 95\% C.L. corresponding to a $\chi^{2}$ value of 7.815. Fig. \ref{fig:2d-oot} shows the correlated 95\% C.L. $\chi^2$ sensitivity contours for $\mathcal{O}_{LHD1}$ in combination with each of the other seven operators. The contours are grouped into different planes based on the relative sensitivities. Stronger correlations are observed between $\mathcal{O}_{LHD1}$ and $\mathcal{O}_{LHD2}$, as well as $\mathcal{O}_{LHDe}$, driven by the larger separation in their sensitivities. The two-parameter limits are consistent with the one-parameter results, which we verify by examining the bounds on one operator when the other is set to zero within each contour of Fig. \ref{fig:2d-oot}. Additional correlated 95\% C.L. contours for all operator pairs are provided in Fig. \ref{fig:2d-oot1} in Appendix \ref{sec:appB}.

\section{Comparison with FCC projections}
\label{sec:fcccomp}
In this section, we present a comparison between the projected sensitivities on the LNV operators from $W^+W^+/,W^+qq'$ production at $\mu$TRISTAN and the best available projections from the phenomenological study \cite{Fridell:2023rtr} involving $\mu^+\mu^{+}jj$ production at the FCC-hh \cite{Benedikt:2022kan,Suarez:2022pcn}, a proposed $pp$ collider operating at a CM energy of 100 TeV with an expected integrated luminosity of up to 30 ab$^{-1}$. The FCC-hh projections in \cite{Fridell:2023rtr} are obtained by scaling background event counts from the ATLAS analysis \cite{ATLAS:2018dcj} to match the 100 TeV, 30 ab$^{-1}$ conditions, and extracting exclusion limits based on the statistical significance of the signal over background.

Tab. \ref{tab:fcc_wc} compares the 95\% C.L. exclusion limits on the WCs, $C_i/\Lambda^3$, and on the effective scale, $\Lambda$ (setting $C_i=1$), between the FCC-hh and our $\mu$TRISTAN 2 TeV, 1 ab$^{-1}$ study. The improvement for operators of the classes $\Psi^2 H^2 D^2$ and $\Psi^2 H^3 D$, contributing to $W^+W^+$ production, is substantial: sensitivities on $\mathcal{O}_{LHD2}$ and $\mathcal{O}_{LHDe}$ improve by about $\mathcal{O}(10^3)$ and $\mathcal{O}(10^2)$, respectively, while $\mathcal{O}_{LHD1}$ improves by roughly an order of magnitude.

For the four-fermion operators contributing via the $W^+qq'$ channel, the FCC-hh performs well due to the presence of light quarks. The sensitivities from $\mu$TRISTAN are of the same order of magnitude and comparable to the FCC-hh bounds, with the FCC-hh offering slightly better limits for $\mathcal{O}_{LLduD}$, $\mathcal{O}_{LLQdH1}$, and $\mathcal{O}_{LLQuH}$, while $\mu$TRISTAN shows better sensitivity for $\mathcal{O}_{LLQdH2}$ and $\mathcal{O}_{LeudH}$.

Overall, this comparison highlights that despite operating at 50 times lower CM energy and 30 times lower integrated luminosity, $\mu$TRISTAN achieves significantly stronger sensitivity particularly for $\Psi^2 H^2 D^2$ and $\Psi^2 H^3 D$ operators and maintains competitive sensitivity for the four-fermion operators compared to the proposed FCC-hh.

\begin{table*}[htb!]
    \centering
    \renewcommand{\arraystretch}{1.2}{
    \begin{tabular}{|>{\centering\arraybackslash}p{3cm}|>{\centering\arraybackslash}p{2.5cm}|>{\centering\arraybackslash}p{3cm}|>{\centering\arraybackslash}p{2.5cm}|>{\centering\arraybackslash}p{3cm}|}
    \hline
        & \multicolumn{2}{c|}{FCC-hh 100 TeV 30 ab$^{-1}$} & \multicolumn{2}{c|}{$\mu$TRISTAN 2 TeV 1 ab$^{-1}$} \\ \cline{2-5} 
        \multirow{-2}*{Operators} & $\Lambda$ (TeV) & $C_i/\Lambda^3$ (TeV$^{-3}$) & $\Lambda$ (TeV) & $C_i/\Lambda^3$ (TeV$^{-3}$)\\
        \hline
        $\mathcal{O}_{LHD1}$ & 4.90 & $8.500 \times 10^{-3}$ & 8.50 & $1.627 \times 10^{-3}$ \\
        $\mathcal{O}_{LHD2}$ & 0.18 & $1.715 \times 10^{2}$ & 1.86 & $1.563 \times 10^{-1}$ \\
        $\mathcal{O}_{LHDe}$ & 0.44 & $1.174 \times 10^{1}$ & 1.88 & $1.511 \times 10^{-1}$ \\
        $\mathcal{O}_{LLduD}$ & 19.0 & $1.458 \times 10^{-4}$ & 14.2 & $3.507 \times 10^{-4}$ \\
        $\mathcal{O}_{LLQdH1}$ & 4.30 & $1.258 \times 10^{-2}$ & 4.02 & $1.534 \times 10^{-2}$ \\ 
        $\mathcal{O}_{LLQdH2}$ & 3.10 & $3.357 \times 10^{-2}$ & 4.04 & $1.510 \times 10^{-2}$ \\
        $\mathcal{O}_{LLQuH}$ & 5.40 & $6.351 \times 10^{-3}$ & 4.92 & $8.451 \times 10^{-3}$ \\
        $\mathcal{O}_{LeudH}$ & 4.50 & $1.097 \times 10^{-2}$ & 4.69 & $9.693 \times 10^{-3}$ \\
        \hline      
    \end{tabular}}
    \caption{Comparison between the projected lower limits on $\Lambda$ and upper limits of WCs for dimension 7 SMEFT operators at the FCC-hh 100 TeV 30 ab$^{-1}$ \cite{Fridell:2023rtr} and $\mu$TRISTAN with $\sqrt{s} $=2 TeV and 1 ab$^{-1}$ luminosity.}
    \label{tab:fcc_wc}
\end{table*}

{
\section{New Physics Implications}
\label{sec:nptoeft}
In this section, we present a concrete UV complete scenario that generates a dimension-7 operator at the leading order, while avoiding the generation of the dimension-5 Weinberg operator at the same order. The construction of such UV completions and their matching onto effective operators has been explored extensively in the literature \cite{Fridell:2024pmw,Bonnet:2009ej,Angel:2012ug,Cai:2014kra,Cepedello:2017eqf,Cepedello:2017lyo,Anamiati:2018cuq,DeGouvea:2019wnq,Gargalionis:2020xvt}. We consider a model involving two heavy scalar leptoquarks (LQs) which, upon integrating out the heavy degrees of freedom, induces the operator $\mathcal{O}_{LLQdH1}$. Although this UV realization has been studied previously \cite{Fridell:2024pmw,Fridell:2024zla,Dorsner:2016wpm,Dorsner:2017wwn,AristizabalSierra:2007nf}, we revisit it here in light of the projected sensitivity of the $\mu$TRISTAN, and derive updated constraints on the model parameters using our collider analysis.

We extend the SM by introducing two scalar LQs, 
$S(\overline{3},1,1/3)$ and $R(3,2,1/6)$, where the quantum numbers denote their representations under the SM gauge group 
$SU(3)_C \times SU(2)_L \times U(1)_Y$. 
For simplicity and phenomenological relevance, we restrict the LQ couplings to the second generation of leptons and the first generation of quarks. The relevant LQ Lagrangian is given by
\begin{equation}
\begin{split}
    \mathcal{L}_{\rm LQ} \supset &\;
    \mathcal{L}^{S}_{\rm kinetic} + \mathcal{L}^{R}_{\rm kinetic} - g_{r}\,(\overline{L}\, i\sigma_{2} R^{\dagger} d) \\
    -&\; g_{s}\,(\overline{Q^{c}}\, i\sigma_{2} S L) - \mu_{rs}\,(S H^{\dagger} R)
    + \text{h.c.}\,,
\end{split}
\end{equation}
where $\sigma_{2}$ denotes the second Pauli matrix, and 
$\mathcal{L}^{S}_{\rm kinetic}$ and $\mathcal{L}^{R}_{\rm kinetic}$ represent the kinetic terms for the $S$ and $R$ fields, respectively. In this setup, neutrino masses are generated radiatively at the one-loop level with quarks propagating in the loop, thereby avoiding the generation of the dimension-5 Weinberg operator at leading order. Integrating out the heavy leptoquark fields induces the dimension-7 operator $\mathcal{O}_{LLQdH1}$, see Fig.~\ref{fig:match}.
\begin{figure}[htb!]
\centering
\begin{tikzpicture}[baseline={(current bounding box.center)},style={scale=0.75, transform shape}]
\begin{feynman}
\vertex(a);
\vertex[above left =1.25cm and 1.25cm of a](a1){$L$};
\vertex[below left =1.25cm and 1.25cm of a](a2){$Q$};
\vertex[right =1.00cm of a](b);
\vertex[right =1.00cm of b](c);
\vertex[above =1.5cm of b](b1){$H$};
\vertex[above right =1.25cm and 1.25cm of c](c1){$L$};
\vertex[below right =1.25cm and 1.25cm of c](c2){$d$};
\diagram*{
(a1) -- [ fermion, arrow size=0.7pt, thick] (a),
(a2) -- [ fermion, arrow size=0.7pt, thick] (a),
(a) -- [ charged scalar, arrow size=0.7pt, thick, edge label' = $S$] (b),
(c) -- [ charged scalar, arrow size=0.7pt, thick, edge label = $R$] (b),
(b) -- [ scalar, arrow size=0.7pt, thick] (b1),
(c1) -- [ fermion, arrow size=0.7pt, thick] (c),
(c) -- [ fermion, arrow size=0.7pt, thick] (c2)
};
\end{feynman}
\end{tikzpicture}
\qquad
\begin{tikzpicture}[baseline={(current bounding box.center)},style={scale=0.75, transform shape}]
\begin{feynman}
\vertex(a);
\vertex[above left =1.25cm and 1.25cm of a] (a1){$L$};
\vertex[below left =1.25cm and 1.25cm of a] (a2){$Q$};
\vertex[above right =1.375cm and 1.375cm of a] (b1){$L$};
\vertex[above =1.5cm of a] (b3){$H$};
\vertex[below right =1.375cm and 1.375cm of a] (b2){$d$};
\diagram*{
(a1) -- [ fermion, arrow size=0.7pt, thick] (a),
(a2) -- [ fermion, arrow size=0.7pt, thick] (a),
(a) -- [ scalar, arrow size=0.7pt, thick] (b3),
(b1) -- [ fermion, arrow size=0.7pt, thick] (a),
(a) -- [ fermion, arrow size=0.7pt, thick] (b2)};
\end{feynman}
\node at (a)[black,fill,style=black,inner sep=3pt]{};
\end{tikzpicture}
\caption{Feynman diagrams illustrating the UV-complete scenario (\textit{left}) and the corresponding matched EFT process
 (\textit{right}). The black square denotes the effective vertex generated by the dimension-seven SMEFT operator $\mathcal{O}_{LLQdH1}$.}
\label{fig:match}
\end{figure}

Matching the UV theory onto the SMEFT yields
\begin{equation}
    \left| \frac{C_{LLQdH1}}{\Lambda^{3}} \right|
    \simeq \frac{\mu_{rs}\, g_{r}\, g_{s}}{m_{S}^{2}\, m_{R}^{2}}\,,
\end{equation}
where $m_{S}$ and $m_{R}$ denote the masses of the $S$ and $R$ LQs, respectively. Using the matching relation above, we translate the projected bounds on the effective operator into indirect constraints on the leptoquark masses. In particular, we infer limits from the sensitivities obtained at the LHC and the proposed $\mu$TRISTAN. For the operator $\mathcal{O}_{LLQdH1}$, the sensitivities of the $\mu$TRISTAN and the FCC-hh are found to be comparable; therefore, we present only the $\mu$TRISTAN limits in the following analysis. To illustrate the impact of these bounds, we consider three representative benchmark scenarios, defined by
\begin{equation}
\begin{split}
    \text{BP1:}\;&\quad \sqrt{g_{r}\,g_{s}} = 1.0\,, \quad \mu_{rs} = 1 \text{ TeV}\,,\\
    \text{BP2:}\;&\quad \sqrt{g_{r}\,g_{s}} = 0.5\,, \quad \mu_{rs} = 1 \text{ TeV}\,,\\
    \text{BP3:}\;&\quad \sqrt{g_{r}\,g_{s}} = 0.5\,, \quad \mu_{rs} = 2 \text{ TeV}\,.
\end{split}
\end{equation}
\begin{figure}[htb!]
    \centering
    \includegraphics[width=\linewidth]{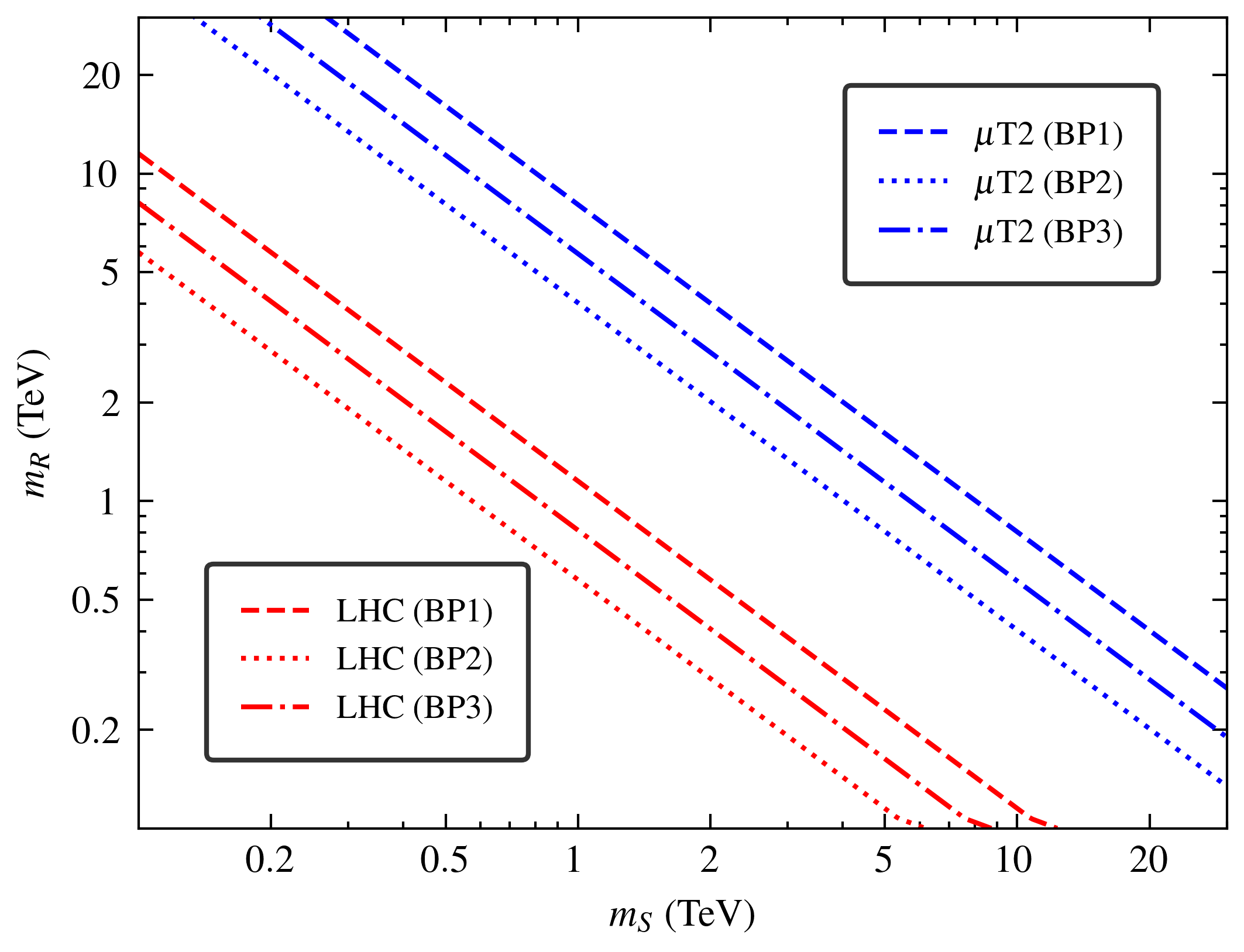}
    \caption{Projected indirect 95\% C.L. constraints on the LQ mass parameter space in the $m_{S}-m_{R}$ plane for different benchmark points (see text), derived from the sensitivities to $C_{LLQdH1}/\Lambda^{3}$ at the LHC (13 TeV 139 fb$^{-1}$) and the $\mu$TRISTAN ($\mu$T2, 2 TeV 1 ab$^{-1}$).}
    \label{fig:eft2}
\end{figure}
Figure \ref{fig:eft2} shows the projected exclusion contours in the $m_{S}-m_{R}$ parameter space obtained by translating the bounds on $(C_{LLQdH1}/\Lambda^{3})$ from the LHC and the $\mu$TRISTAN. The results indicate that the indirect constraints on the LQ masses are strengthened by approximately an order of magnitude at the $\mu$TRISTAN (2 TeV 1 ab$^{-1}$) compared to those achievable at the LHC (13 TeV 139 fb$^{-1}$), highlighting the potential of the $\mu$TRISTAN to significantly enhance indirect sensitivity to heavy BSM states.
}
\section{Summary and Conclusion}
\label{sec:conclusion}
In this work, we investigated the prospects of probing dimension-7 $\Delta L=2$ operators at the future same-sign muon collider $\mu$TRISTAN, which is expected to operate at a center-of-mass energy of 2 TeV with a projected integrated luminosity of 1 ab$^{-1}$. Having same sign di-muon at the initial stage, any final state without leptons can potentially 
indicate to LNV processes in action and can provide better sensitivity than the presently running or even future sensitivities of the LHC. 

The operators that contribute to such signal can be classified mainly into four categories: $\Psi^{2} H^{2} D^{2}$, $\Psi^2 H^3 D$, 
$\Psi^4 D$, and $\Psi^4 H$. We mainly focus on muon flavoured operators, as we probe them at $\mu$TRISTAN, having lighter constraints from 
observables like neutrino less double beta decay. Note here that dimension five Weinberg operator also provides lepton number violation, but connects 
only to neutrinos, so that it is difficult to connect them to observables at the collider. Our analysis is focused on $W^+W^+$ and $W^+qq'$ production, 
with the final-state signature comprising two fat jets. Operators that produce the signal via contact interaction, provides larger cross-section, than those having propagator suppression. 

The SM background mainly arises from $W^+W^+\bar{\nu}\bar{\nu}$ process. Event selection criteria and kinematic cuts were carefully designed to effectively suppress SM backgrounds, with particular emphasis on enhancing sensitivity to different classes of LNV operators. Missing energy turns out to be one of the most useful variables in this 
process apart from the invariant mass between the two fat jets. All the operators easily surpass the discovery limit at the $\mu$TRISTAN, for the Wilson coefficients $C_i/\Lambda^3$ consistent with present experimental constraints. 

We also performed a binned $\chi^2$ analysis based on the angular distribution of the heavier fat jet to extract projected exclusion sensitivities on the $C_i/\Lambda^3$ of the LNV operators. Both individual operator and correlated two-operator scenarios were explored, deriving projected 68\% and 95\% C.L. exclusion limits. The projected limits surpass the existing LHC bounds for each operator by at least an order of magnitude. Furthermore, we compared our results with projections from a recent phenomenological study at the FCC-hh 100 TeV 30 ab$^{-1}$ run. Despite the $\mu$TRISTAN operating at a significantly lower center-of-mass energy and luminosity, it achieves substantially improved sensitivities for operators of the $\Psi^2 H^2 D^2$ and $\Psi^2 H^3 D$ classes and competitive sensitivities for the four-fermion operators contributing via the $W^+qq'$ channel. Our study highlights the unique potential of same-sign muon colliders like $\mu$TRISTAN to probe LNV physics with clean final states and precise kinematic control, complementing and, in some cases, surpassing the reach of future hadron colliders.

\begin{acknowledgments}
SB would like to acknowledge DST ANRF Grant CRG/2023/000580 from the Govt. of India.
\end{acknowledgments}

\appendix
{
\section{The Weinberg operator}
\label{sec:appA}
In this section, we present the projected sensitivity of $\mu$TRISTAN to the Weinberg operator specific to 
the muon-flavor–diagonal case, and compare its reach with that of the current LHC and the proposed FCC-hh. Direct neutrino mass measurements based on muon kinematics \cite{Assamagan:1995wb} constrain the corresponding Majorana mass parameter to be $m_{\mu\mu} < 0.17$ MeV. This translates into a muon-flavor–diagonal NP scale of order $\Lambda_{\mu\mu} \sim 10^{5}$~TeV. While this bound is weaker than the one obtained for the electron sector, it nevertheless lies far beyond the direct reach of present-day colliders. {It is worth noting that current cosmological observations constrain the sum of the light neutrino masses to $\sum m_\nu \lesssim 0.12$ eV (95\% C.L.)~\cite{Planck:2018vyg}, which is considerably more stringent than existing direct experimental bounds.}

The sensitivity of hadron colliders to the Weinberg operator is provided for dimension-seven operator $\mathcal{O}_{LH}$ in \cite{Fridell:2023rtr}. The effective scale $\Lambda$ associated with $\mathcal{O}_{LH}$ can be related to the muon-diagonal NP scale $\Lambda_{\mu\mu}$ via
\begin{equation}
    \Lambda_{\mu\mu} = \left(\frac{2\Lambda^{3}}{v^{2}}\right),
\end{equation}
where $v$ denotes the Higgs vacuum expectation value. Using the existing constraints on $\mathcal{O}_{LH}$ from the LHC at $\sqrt{s}=13$~TeV with an integrated luminosity of $139~\text{fb}^{-1}$, as well as the projected sensitivities for the FCC-hh at $\sqrt{s}=100$~TeV with $30~\text{ab}^{-1}$ \cite{Fridell:2023rtr}, we obtain the following bounds:
\begin{equation}
\begin{split}
    \Lambda_{\mu\mu} &> 1.42~\text{TeV} \quad (\text{LHC})\,,\\
    \Lambda_{\mu\mu} &> 21.8~\text{TeV} \quad (\text{FCC})\,.
\end{split}
\end{equation}

At $\mu$TRISTAN, the Weinberg operator can be probed through $t$-channel same-sign $W^{+}W^{+}$ production, mediated by neutrino fusion via the Majorana mass insertion $\overline{\nu^{c}}\nu$. We focus on the two fat jet final state and apply the signal selection and kinematic cuts described in Sec. \ref{sec:collider}, resulting in an overall signal efficiency of $0.88$. Repeating the $\chi^{2}$-based statistical analysis outlined in Sec. \ref{sec:constraints}, we derive the following limits on the NP scale:
\begin{equation}
\begin{split}
    \Lambda_{\mu\mu} &> 476~\text{TeV} \quad (68\%~\text{C.L.})\,, \\
    \Lambda_{\mu\mu} &> 399~\text{TeV} \quad (95\%~\text{C.L.})\,.
\end{split}
\end{equation}

Although these bounds remain weaker than those inferred from direct neutrino mass measurements, they surpass the projected FCC-hh sensitivity by more than an order of magnitude, highlighting the exceptional potential of $\mu$TRISTAN in probing lepton number violating physics in the muon sector.

\section{Validity of the effective field theory}
\label{sec:appX}
The EFT description of an NP scenario relies on a perturbative expansion of the scattering amplitude in inverse powers of the NP scale, $\Lambda$, assuming Wilson coefficients of $\mathcal{O}(1)$. At the amplitude level, the EFT contribution can be schematically written as
\begin{equation}
\mathcal{A} \;=\; \sum_{d} \, \widetilde{\mathcal{A}}_{d}
\left(\frac{p}{\Lambda}\right)^{d-4},
\end{equation}
where $d$ denotes the canonical dimension of the effective operator and $p$ represents the characteristic momentum scale of the process. The perturbative expansion is well defined only when the expansion parameter is sufficiently small, i.e.
\begin{equation}
p \ll \Lambda,
\end{equation}
which ensures convergence of the series and the validity of the EFT description~\cite{Manohar:2018aog}.

A commonly adopted criterion for EFT validity is that the cutoff scale lies above the largest energy scale probed in the process; at collider experiments, this implies that the accessible momentum transfer remains below $\Lambda$. For lepton colliders, the maximum momentum transfer is set by the CM energy, $p \sim \sqrt{s}$, leading to the condition
\begin{equation}
\label{eq:a2a}
\sqrt{s} < \Lambda \,.
\end{equation}
In the case of $\mu$TRISTAN, where we consider $\sqrt{s}=2$ TeV, the EFT analysis is therefore self-consistent only for NP scales satisfying $\Lambda > 2$ TeV. This ensures that contributions from higher-dimensional operators remain perturbative and that the underlying UV dynamics remain effectively decoupled.

However this criterion provides only an approximate assessment of EFT validity; a rigorous determination requires knowledge of the explicit UV completion from which the effective operators originate. The EFT approach is justified only when the available energy is insufficient to produce the heavy degrees of freedom on-shell requiring
\begin{equation}
\sqrt{s} < m\,,
\end{equation}
where $m$ denotes the mass of the heavy NP mediator. In the EFT framework, the combination $C/\Lambda^{d-4}$ encodes both the propagator suppression due to the heavy mass and the underlying couplings, rendering the condition in Eq.~\eqref{eq:a2a} only approximate.

To illustrate this explicitly, we examine the dependence of the EFT cutoff on the parameters of the UV complete model introduced in Sec. \ref{sec:nptoeft}. Fixing the LQ masses to $M(=m_{S}=m_{R})=2$ TeV, we study the variation of the inferred EFT scale $\Lambda$ as a function of the couplings $\sqrt{g_{r}g_{s}}$ and the trilinear parameter $\mu_{rs}$. The results are shown in Fig. \ref{fig:eft1}. We observe that the assumption $\Lambda > 2$ TeV is satisfied only for specific regions of the BSM parameter space, highlighting the model-dependent nature of EFT validity.
\begin{figure}[htb!]
\centering
\includegraphics[width=\linewidth]{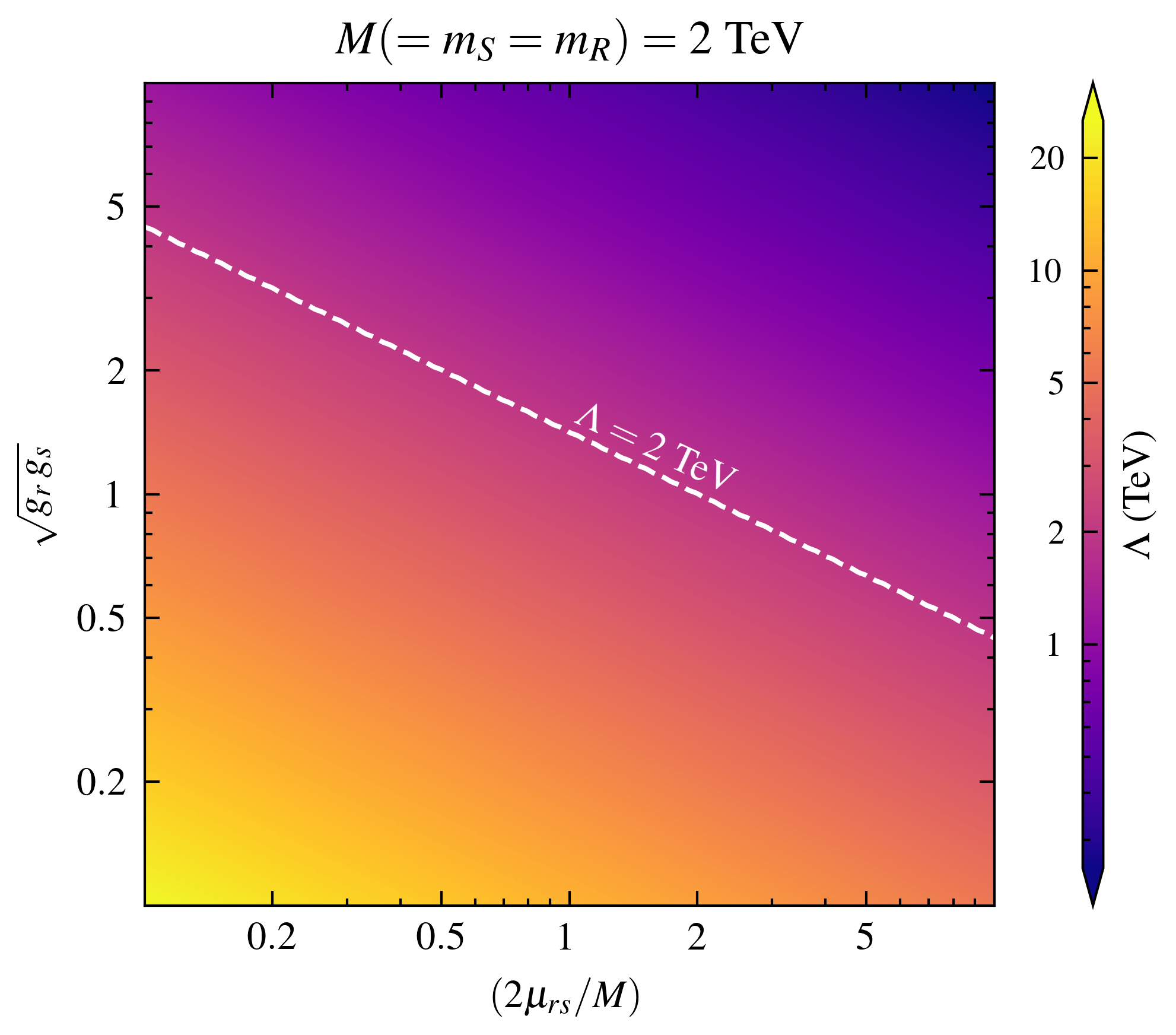}
\caption{Variation of the inferred EFT cutoff scale $\Lambda$ (shown by the color bar) in the $\mu_{rs}-\sqrt{g_{r}g_{s}}$ parameter space. The LQ masses are fixed to $2$ TeV.}
\label{fig:eft1}
\end{figure}
}
\section{Additional sensitivity plots}
\label{sec:appB}
\begin{figure*}[htb!]
    \centering
    \includegraphics[width=\linewidth]{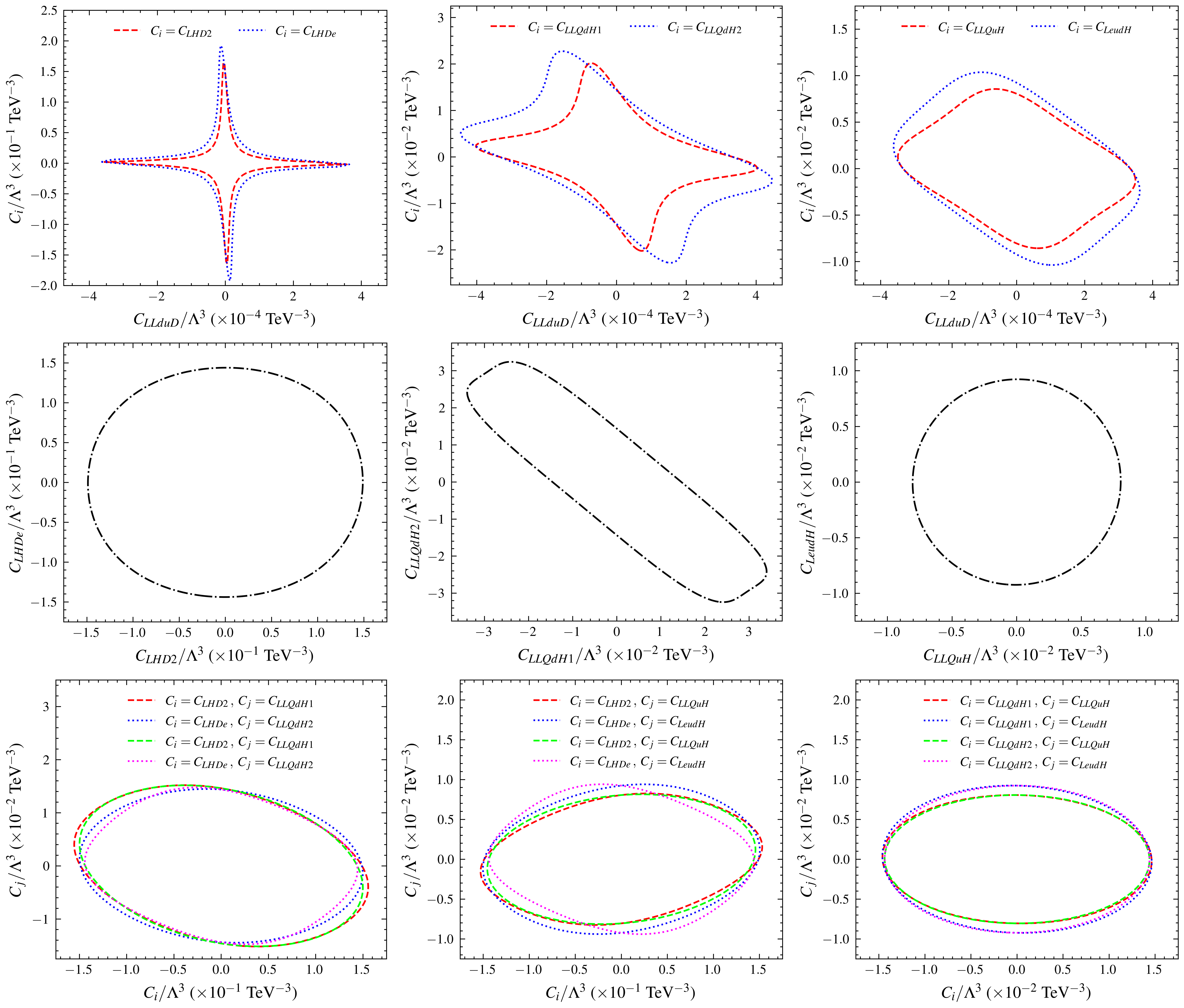}
    \caption{95\% C.L. $\chi^{2}$ sensitivity contours for different LNV operator combinations, derived from $W^+W^+/\;W^+qq'$ production at the $\mu$TRISTAN with $\sqrt{s} = 2$ TeV and an integrated luminosity of $\mathfrak{L}_{\rm int} = 1$ ab$^{-1}$.}
    \label{fig:2d-oot1}
\end{figure*}

The 95\% C.L. $\chi^{2}$ sensitivity contours for various operators (excluding $\mathcal{O}_{LHD1}$) are shown in Fig. \ref{fig:2d-oot1}, plotted in the parameter space of the WCs of the operators. The corresponding WCs are indicated in the plot labels for single-curve plots or in the legends for multi-curve plots.

\bibliography{d7ssmc}
\end{document}